\documentclass[conference]{IEEEtran}

\usepackage[T1]{fontenc}
\usepackage{hhline}
\usepackage{amssymb,amsfonts,amsthm}
\usepackage{amsmath}
\usepackage{epsfig}
\usepackage{color}
\usepackage{enumerate}
\usepackage{ifpdf}
\usepackage{graphicx}
\usepackage[ruled,vlined,linesnumbered]{algorithm2e}
\usepackage[tight,footnotesize]{subfigure}
\usepackage{cite}

\begin{document}

\title{Network Coding Function Virtualization}

\author{Tan Do-Duy, M. Angeles Vazquez Castro\\
Dept. of Telecommunications and Systems Engineering\\
Autonomous University of Barcelona, Spain\\
Email: \{tan.doduy, angeles.vazquez\}@uab.es}

\maketitle
\begin{abstract}
Network Functions Virtualization (NFV) and Network Coding (NC) have
attracted much attention in recent years as key concepts for providing
5G networks with flexibility and differentiated reliability, respectively.
In this paper, we present the integration of NC architectural design
and NFV. In order to do so we first describe what we call a virtualization
process upon our proposed architectural design of NC that should help
to offer the reliability functionality to a network. The process consists
of identifying the required functional entities of NC and analyzing
when the functionality should be activated towards complexity/energy
efficiency. The relevance of our proposed NC function virtualization
is its applicability to any underlying physical network, satellite
or hybrid thus enabling softwarization, and rapid innovative deployment.
Finally, we validate our framework to a study case of geo-control
of network reliability that is based on device's geographical location-based
signal/network information.
\end{abstract}

\section{Introduction \label{sec:intro}}

NC has recently emerged as a new approach for improving network performance
in terms of throughput and reliability, especially in wireless networks.
Coding operations are executed at the source and/or re-encoded at
intermediate nodes so that receivers are able to recover from losses
at the receiver side.

NFV has been proposed as a promising way the telecommunications sector
facilitates the design, deployment, and management of networking services.
Essentially, NFV separates software from physical hardware so that
a network service can be implemented as a set of virtualized network
functions through virtualization techniques and run on commodity standard
hardware. NFV can be instantiated on demand at different network locations
without requiring the installation of new equipment. This key idea
helps network operators deploy new network services faster over the
same physical hardware which reduces capital investment and the time
to market of a new service. Now it is the turn for the telecommunications
technologies to detach software from the hardware where the telecommunications
functions run (which makes equipment seller-dependent and expensive).
Such detachment demands functionalities and infrastructures to be
designed under such new paradigm.

The integration of NC and virtualization opens applicability of NC
functionalities in future networks (e.g. upcoming 5G networks) to
both distributed (i.e. each network device) and centralized manners
(i.e. servers or service providers). The European Telecommunications
Standards Institute (ETSI) has proposed some use cases for NFV in
\cite{NFV001.2013}, including the virtualization of cellular base
station, mobile core network, home network, and fixed access network,
etc. In particular, there are already available some other efforts
in the combination of NC virtualization and SDN technology. For example,
in \cite{Szabo2.2015}, the authors investigate the usability of random
linear NC as a function in SDNs to be deployed with virtual (software)
OpenFlow switches. The NC functions including encoder, re-encoder,
and decoder are run on a virtual machine (ClickOS). This work provided
a prototype of combination of NC and SDN. It also includes the implementation
of additional network functions via virtual machines (VMs) by sharing
system resources or additional hardware (e.g. FPGA cards). In \cite{Hansen.2015},
integration of network coding run in a VM into an Open vSwitch shows
the relation between coding software, VM and host OS. The paper indicated
the integration of NC as a functionality of a software defined network
is possible. However, a unified theoretical framework for NC design
in view of NFV is currently missing.

The contributions of this paper can be summarized as follows:
\begin{itemize}
\item An architectural design framework for NC is proposed, which includes
a functional domain design that enables virtualization-based design.
\item A virtualization process is presented by which NC can operate as a
reliability functionality to any physical underlying network.
\item We validate our framework for an illustrative study case of geo-control
network reliability in which the geographical location-based information
of network devices has been exploited in the definition of the required
functional entities.
\end{itemize}
The rest of this paper is organized as follows. In Section \ref{sec:NCframework},
we propose the architectural design of NC. In Section \ref{sec:Virtualization},
we present higher-level architecture and virtualization process identifying
an instance of NC functionalities to the network. In Section \ref{sec:Reliability},
we show when NC functionality is activated considering complexity/energy
efficiency and also analyze the feasibility of the ideas with today's
technologies via a use case of NCFV for geo-controlled networks. Finally,
in Section \ref{sec:CONC}, we identify conclusion and further work.

\section{NC Architectural Design Domains \label{sec:NCframework}}

Virtualization and NC are two different techniques to address different
challenges in the designs of upcoming network technologies. The combination
of virtualization and NC brings forth a potential solution for the
management and operation of the future networks \cite{Hansen.2015}.
In order to do so, we present a proposal of distinct design domains.
In order to distinguish the different abstraction processes: the ones
related to NC and the ones related to NCFV.

Such domains are proposed as follows \cite{MAVazquez.2015}: 
\begin{itemize}
\item \textbf{NC coding domain} - domain for the design of coding theoretical
network codes, encoding/decoding algorithms, performance benchmarks,
appropriate mathematical-to-logic maps, etc.
\item \textbf{NC functional domain} - domain for the design of the functional
properties of NC to match design requirements built upon abstractions
of 

\begin{itemize}
\item \textbf{Network}: by choosing a reference layer in the standardized
protocol stacks and logical nodes for NC and re-encoding operations. 
\item \textbf{System}: by abstracting the underlying physical or functional
system at the selected layer e.g. SDN and/or function virtualization. 
\end{itemize}
\item \textbf{NC protocol domain} - domain for the design of physical signaling/transporting
of the information flow across the virtualized physical networks in
one way or interactive protocols. 
\end{itemize}
The domain relevant for NC as NFV is NC functional domain and we develop
this domain in the next section by first interpreting NC as a network
service for control of reliability. It is noted that use cases of
NFV mentioned (e.g. \cite{NFV001.2013}) investigate the concentration
of network functions in centralized architectures such as data centers
or centralized locations proposed by network operators, service integrators
and providers. Nevertheless, our proposal is of distributed nature,
closer to Distributed-NFV (D-NFV) \cite{RAD.2014}, which provides
distribution of NFV throughout the network. Thus NCFV can be located
at various parts of the network including both centralized locations
and distributed network nodes.

\section{NC Function Virtualization \label{sec:Virtualization} }

\subsection{NC as a Network Functionality \label{sec:Service}}

\begin{figure}[htbp]
\begin{centering}
\includegraphics[scale=0.29]{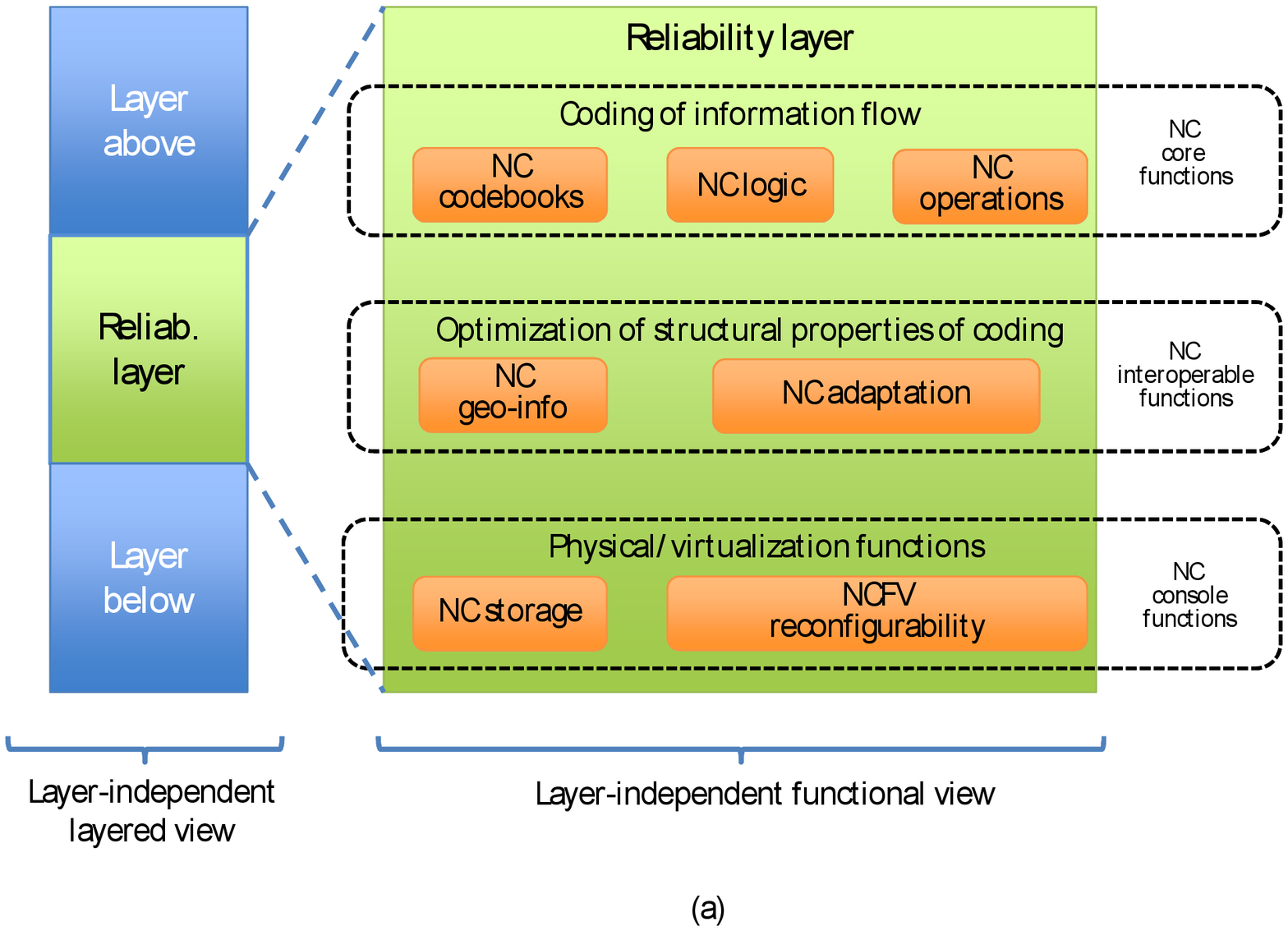}
\par\end{centering}
\begin{centering}
\includegraphics[scale=0.55]{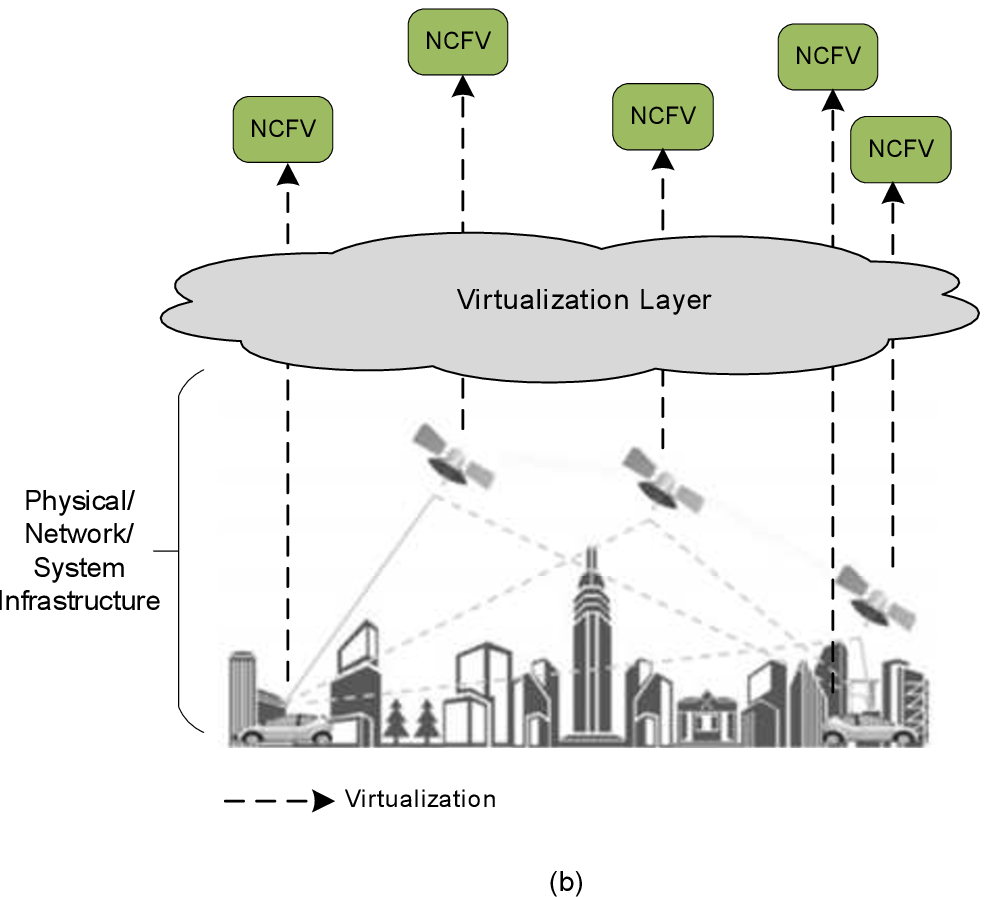}
\par\end{centering}
\caption{(a) Layer-independent functional view of reliability layer and (b)
graph presentation denotes higher-level architecture of NCFV in which
NC is a service of the network.\label{fig:NC-as-Service}}
\end{figure}

In Fig. \ref{fig:NC-as-Service}(a), we illustrate the layer-independent
functional view of a reliability layer at which NC functionality would
operate which we will identify in detail in Section \ref{sec:Process}.
Assume that the NC functionalities have been designed well. Then,
in view of higher-level architecture of NCFV, Fig. \ref{fig:NC-as-Service}(b)
shows that the small boxes in green work according to the NC functionalities
we have identified while at the same time the overall design is compliant
with existing proposals of NFV architectures such as \cite{NFV002.2013}.
In terms of D-NFV, NCFV may be resided anywhere in the network, e.g.
data centers, central offices, and mobile devices, etc.

The physical/network/system infrastructure, e.g. satellite networks
or terrestrial wireless systems, represents the physical resources
including computing, storage, and network that provide processing,
storage, and connectivity to NCFV through virtualization layer respectively.
The difference will be that every virtual infrastructure will have
its corresponding time/space scales and communication/computation
resources.

In the architectural view, the virtualization layer makes sure that
the NCFV is separated from hardware resources so that the softwarized
function can be deployed on different physical hardware resources.
Typical solutions for the deployment of NFV are hypervisors and VMs.
Furthermore, NFV can be realized as an application software running
on top of a non-virtualized server by means of an operating system
(OS) \cite{NFV002.2013}.

\subsection{Virtualization Process \label{sec:Process}}

Based on the key features of NFV design, we identify preliminary NC
functionalities so that NC can provide the reliability as a service
to network (see Fig. \ref{fig:NC-as-Service}(a)). In particular,
we define common functionalities that NC design needs for system-specific
designs \cite{MAVazquez2.2015} as follows:
\begin{itemize}
\item \textbf{NC core functions} 

\begin{itemize}
\item NC Logic: logical interpretation of coding use, coding scheme selection
(intra-session/inter-session, coherent/incoherent, file-transfer/streaming,
systematic/non-systematic), coding coefficients selection (random/deterministic),
etc. 
\item NC Coding: elementary encoding/re-encoding/decoding operations, encapsulation/de-encapsulation,
adding/removing headers, etc. 
\end{itemize}
\item \textbf{NC interoperable functions} 

\begin{itemize}
\item NC Resource Allocation/Adaptation: optimal allocation of NC parameters. 
\item NC Congestion Control: controlling congestion, interoperable with
other congestion-control algorithms. 
\end{itemize}
\item \textbf{NC console functions} 

\begin{itemize}
\item NC Storage: interactions with physical memory. 
\item NC Feedback Manager: settings for feedback. 
\item NC Signaling: coordinating signaling parameters. 
\end{itemize}
\end{itemize}
At the NC core blocks, interactions with other nodes bring into agreement
on coding schemes, coefficients selection, etc. NC coding block receives
all inputs regarding coding scheme, coefficients, coding parameters,
packets from storage block, and signaling to perform elementary encoding/re-encoding/decoding
operations, etc.

At the NC interoperable blocks, NC adaptation block generates optimal
coding parameters to the coding operation block. The NC interoperable
functions also contain NC geo-information block which provides geographical
location-based information and required level of reliability to the
resource allocation process.

At the last stage, NC console blocks, which connect directly to physical
storage and feedback from network nodes to provide NC packets and
packet loss to the upper stage respectively. Depending on every specific
design, the NC function domain blocks allow the designer to adjust
NC functionalities according to the technical requirements and NC
design objectives.

It is noted that in NFV, a virtualized NC function is a software package
that implements such network function. Our proposed functionalities
allows for different (use-case driven) possible implementations of
such package.

\section{Case Study: Geo-controlled Network Reliability \label{sec:Reliability}}

\subsection{Scenario \label{sec:Scenario}}

\begin{figure}[tbh]
\includegraphics[scale=0.5]{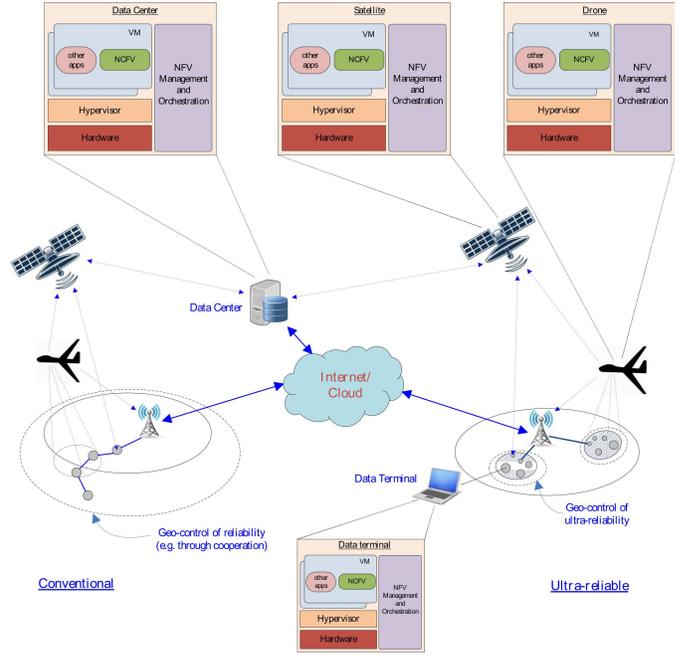}\label{fig:Geo-NCFV}
\caption{Network abstraction showing geo-control of reliability functionality.\label{fig:Geo}}
\end{figure}

Fig. \ref{fig:Geo} illustrates two cases of geo-control of reliability
called conventional and ultra-reliable communication, respectively.
The former case utilizes cooperation with neighboring devices as repeaters
in line networks or multi-hop networks to support communication services
beyond the cell coverage. Whereas the latter case constitutes connections
with surrounding devices to support communication links beyond the
coverage of cellular networks in which NCFV can be located at distributed
network nodes including user terminals, data centers, drone, and satellites
to offer the reliability functionality to the network.

In the following, we are to validate our proposed framework, but is
out of the scope of this paper the actual implementation over VMs.

\subsection{Optimal strategy of virtual reliability function\label{sec:optimized}}

In this section we are interested to validate NCFV operation as a
(virtual) functionality. 

\subsubsection{Analytical model}

Let $\delta$ be the vector of per-hop erasure rates e.g. $\delta=(\delta_{0},\delta_{1})$
for 2 hops and $R=n/m$ be the coding rate and $P_{L}^{NC}(R,\delta)$
be the packet loss rate after decoding. The packet successfully-received
rate after decoding is defined as $P_{R}^{NC}(R,\delta)=1-P_{L}^{NC}(R,\delta)$.
Let $L$ and $s=L/q$ denote packet length in bits and in symbols,
respectively. Let $N^{mult}=(m-n).n.s$ and $N^{add}=(m-n).(n-1).s$
denote the complexity in terms of number of multiplications and additions,
respectively, required for the encoding process. 

We are interested in investigating when NC should be activated in
terms of (1) computational complexity/energy consumption and (2) minimum
requirement of packet delivery rate ($\rho_{0}$). In order to do
so, we define the following utility function: 
\begin{equation}
u^{act}(R,\delta,\rho_{0})=w_{NC}.f^{NC}(R,\delta,\rho_{0})-w_{comp}.f^{comp}(R)
\end{equation}
where the first factor accounts for the goodness of the encoding/decoding
scheme in achieving target performance $\rho_{0}$. The second factor
accounts for the cost in computational complexity/energy consumption.
Whereas $w=(w_{NC},w_{comp})$ with $w_{NC},w_{comp}\in(0,1)$ and
$w_{NC}+w_{comp}=1$ are weigh factors with respect to the goodness
and cost.

We define the goodness and cost, respectively as follows 

\begin{equation}
f^{NC}(R,\delta,\rho_{0})=\frac{P_{R}^{NC}(R,\delta)-\rho_{0}}{\rho_{0}}
\end{equation}

\begin{equation}
f^{comp}(R)=log(1+\bar{\beta}^{enc}(R))
\end{equation}

with $\bar{\beta}^{enc}(R)=\frac{\beta^{enc}(R)}{\beta_{min}^{enc}}$
is the ratio between the encoding complexity and the encoding with
minimum redundancy. We assume that the source has limitation on computational
complexity/energy $\bar{\beta}^{enc}(R)\le\beta_{0}$ (e.g. $\beta_{0}=1.4$).

The source identifies at which rate that it maximizes its own utility
under constraint of complexity/energy by the following optimization
strategy:

\begin{equation}
\underset{R}{argmax}\:u^{act}(R,\delta,\rho_{0},w)
\end{equation}

\[
subject\:to\:\bar{\beta}^{enc}(R)\le\beta_{0}.
\]

This strategy determines the upper bound for the benefit of the source
and the highest performance $P_{R}^{NC}(R,\delta)$, respectively
that the source can provide to the destination according to a given
$(\delta,\rho_{0})$. NC functionality is activated if $u^{act}(R,\delta,\rho_{0})\ge u_{0}$,
where $u_{0}$ is a threshold. Note that given the constraints, our
virtual reliability functionality should self-adapt to underling (physical)
channel and hardware computational limitations.

\subsubsection{Numerical results}

Our proposed optimization strategy Eq. (4) identifies the optimal
points that bring highest benefit for the source's viewpoint. In addition,
it also identifies the upper bound on the achievable performance for
the configuration of the functionality given as $(\delta,\rho_{0},w)$.

\begin{figure}[tbh]
\begin{raggedright}
\includegraphics[scale=0.18]{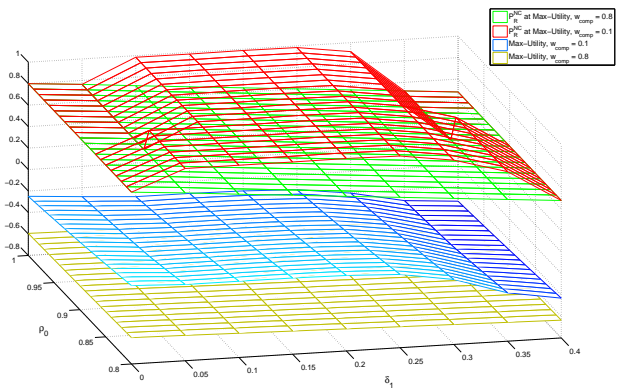}
\par\end{raggedright}
\caption{Maximum utility and respective $P_{R}^{NC}(R,\delta)$ according to
various values of $\rho_{0}$ and $\delta_{1},$ with $w_{comp}=0.1$
and $0.8$.\label{fig:Optimized-strategy}}
\end{figure}

Figures \ref{fig:Optimized-strategy} shows maximum utility and respective
$P_{R}^{NC}(R,\delta)$ for e.g. NC over 2-hop networks \cite{Saxena2.2015}
according to various values of $(\delta_{1},\rho_{0})$ with $\beta_{0}=1.4$,
$\delta_{0}=0.2$, and $\delta_{1}\text{\ensuremath{\in}[0, 0.4]}$.
We consider the optimization problem with small and large values of
$w_{comp}$, respectively. In case of $w_{comp}=0.1$, we can see
that when channel condition is good (i.e. $\delta_{1}$ is small),
the source should activate NC functionality with high coding rate
in order to maximize its own utility with the optimal $P_{R}^{NC}(R,\delta)$
is around 0.8. While NC functionality should be activated with low
coding rate since $\delta_{1}$ increases. Then, $P_{R}^{NC}(R,\delta)$
reaches to 1. In cases of bad channel conditions (e.g. $\delta_{1}=0.4$),
maximum utility and respective $P_{R}^{NC}(R,\delta)$ are very low
due to strong impact of channel conditions. In such cases, the source
should deactivate NC functionality. On the other hand, for $w_{comp}=0.8$
(i.e. high cost in terms of complexity/energy), the results indicate
that maximum utility obtained is very small if compared with the case
of small $w_{comp}$. Moreover, respective $P_{R}^{NC}(R,\delta)$
is small and reduced gradually with the increase of $\delta_{1}$.
Therefore, in these cases, NC functionality should not be activated.

The upper bound on the achievable utility has been identified by the
constraint of complexity/energy. Then, it is necessary to identify
operative ranges of performance so that the destination user is aware
and admits some variations in the quality. The cognitive algorithm
to identify the operative ranges is briefly described as follows:
(1) the source identifies maximum utility ($u_{max}^{act}(R,\delta,\rho_{0},w)$)
and respective $R,\:P_{R}^{NC}(R,\delta)$ given $(\delta,\rho_{0},w)$,
(2) the source then determines $R$ that satisfies $u_{min}^{act}(.)$
$\leq u^{act}(R,\delta,\rho_{0},w)\leq u_{max}^{act}(.)$, where $u_{min}^{act}(.)$
is the lower bound e.g. $u_{min}^{act}(.)=0.8u_{max}^{act}(.)$, and
(3) the source should activate NC functionality if the range of performance
is acceptable by users.

\begin{figure}[tbh]
\begin{centering}
\includegraphics[scale=0.19]{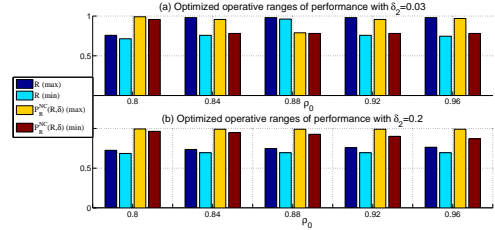}
\par\end{centering}
\caption{Operative ranges of performance with respect to (a) $\delta_{1}=0.03$
and (b) $\delta_{1}=0.2$, with $w_{comp}=0.1$.\label{fig:BarPlot}}
\end{figure}

Fig. \ref{fig:BarPlot} shows the operative ranges of achievable performance
according to $m=50$, $L=100B$, $\delta_{1}=0.03$ and $0.2$, respectively.
We can observe that larger value of $\delta_{1}$ requires smaller
coding rates to obtain highest benefit for the source (i.e. a larger
number of redundant packets). Furthermore, the results show that in
case of large $\delta_{1}$, the range of $P_{R}^{NC}(R,\delta)$
with regard to each case of $\rho_{0}$ is also smaller than that
of small $\delta_{1}$. As a final remark, we note that with larger
$\delta_{1}$, it is required lower coding rates, i.e. higher cost
in terms of computational complexity/energy consumption to reach the
optimal points.

\begin{figure}[tbh]
\includegraphics[scale=0.27]{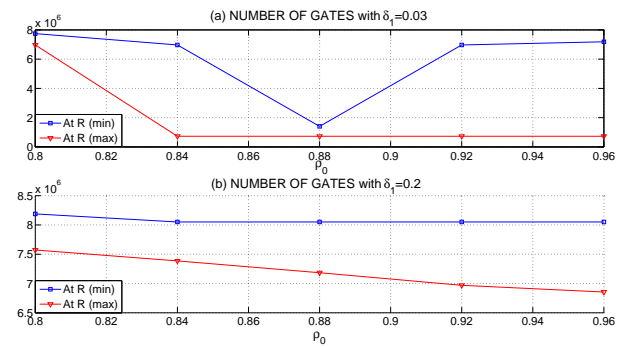}
\caption{Complexity in terms of number of logic gates according to $\delta_{1}=0.03$
and $\delta_{1}=0.2$, respectively.\label{fig:Complexity}}
\end{figure}

In addition, we also evaluate computational complexity in terms of
logic gates. The implementation of addition and multiplication over
$GF(2^{q})$ corresponds to $q$ and $2q^{2}+2q$ gates, respectively
\cite{Angelopoulos.2011}. Figure \ref{fig:Complexity} indicates
the complexity in terms of logic gates with respect to operative ranges
in Figure \ref{fig:BarPlot}. The wider range of the peformance results
in a larger difference between complexity required for minimum and
maximum $P_{R}^{NC}(R,\delta)$ . Therefore, it is apparent to note
that there exists a tradeoff between network performance and computational
complexity.

\subsection{Gain in increase of connectivity\label{sec:Gain}}

The increase in connectivity is defined as the extension of communication
range beyond cellular coverage given a requirement of reliability.
We evaluate the potential of our solution for the increase of connectivity
through simulation results. We assume that packet erasure rates for
all links are $0.03$. Whereas $w_{comp}=0.1$ and $\rho_{0}=0.9$,
respectively. The analysis of $P_{R}^{NC}(R,\delta)$ indicating the
ability of the NC virtualized design in supporting geo-controlled
network reliability is conducted in \cite{Tan.2015}. It is assumed
that a great number of network devices are uniformly distributed in
a deployment area of $6$ $km$ $\times$ $6$ $km$. Then, the ratio
of the deployment area and the number of devices defines network density
i.e. the average area covered by a device. A base station (BS) is
located at center of the area with maximum radio range of 1.5 km.
It is assumed that maximum radio range of WiFi signal on each device
is $50$ $m$. The number of devices is $10200$0 and $150000$, respectively
with 1 device per $350$ $m^{2}$ and $1$ device per $250$ $m^{2}$.
By limited radio range of BS, the deployment area is divided into
two regions: within cellular coverage and beyond cellular coverage.

Using statistics based on some uniform symmetry assumption over the
2D coverage, we identify the number of available neighbors for a given
density. For high uniform density of devices and all having the same
loss rate, we could compute the percentage of available relays around
a device, but also of how many hops each device could have. The same
is for all devices in the deployment area. Whereas for low density
that percentage would decrease. Then, we have a law of how the $P_{R}^{NC}(R,\delta)$
varies with distance for different densities, which corresponds to
a given loss rate and the number of hops. The ratio of the short-range
radio coverage and device's density identifies the number of neighbors
around a device, i.e. relays that may help with improving network
reliability. The distance is normalized by coverage range of cellular
signal to compare how much the coverage is improved beyond cellular
range.

\begin{table}[htbp]
\caption{GAIN IN INCREASE OF CONNECTIVITY WITH/WITHOUT GEO-NC FOR A (HIGH)
DENSITY OF 1 DEVICE PER 250 $m^{2}$.\label{tab:Table3}}

\centering{}%
\begin{tabular}{|c|c|c|c|}
\hline 
Required $P_{R}^{NC}(R,\delta)$ & 95 \% & 90 \%  & 85 \%\tabularnewline
\hline 
\hline 
With Geo-NC  & 27 \%  & 60 \%  & 87.5 \%\tabularnewline
\hline 
Without Geo-NC  & 1.5 \% & 12.5 \%  & 20 \%\tabularnewline
\hline 
Gain  & 18  & 4.8 & 4.3\tabularnewline
\hline 
\end{tabular}
\end{table}

\begin{table}[htbp]
\caption{GAIN IN INCREASE OF CONNECTIVITY WITH/WITHOUT GEO-NC FOR A (LOW) DENSITY
OF 1 DEVICE PER 350 $m^{2}.$\label{tab:Table4}}

\centering{}%
\begin{tabular}{|c|c|c|c|}
\hline 
Required $P_{R}^{NC}(R,\delta)$  & 95 \% & 90 \%  & 85 \%\tabularnewline
\hline 
\hline 
With Geo-NC  & 12 \%  & 19.5 \% & 25.5 \%\tabularnewline
\hline 
Without Geo-NC  & 1 \% & 9.5 \%  & 16.5 \%\tabularnewline
\hline 
Gain & 12 & 2.1 & 1.6\tabularnewline
\hline 
\end{tabular}
\end{table}

Focusing on the performance of geo-control, Tables \ref{tab:Table3}
and \ref{tab:Table4} show $P_{R}^{NC}(R,\delta)$ (i.e. reliability)
for different densities corresponding to a given loss rate. In particular,
we note that with NC functionality, the requirement of $90\%$ can
be reached at normalized distance of $1.6$, i.e. $60\%$ extension
beyond the cellular coverage in case of 1 device per 250 $m^{2}$.
The reason is that spatial diversity created by up to 4 relays increases
opportunity for the network-coded packets to reach the destinations.
Otherwise, for low device's density e.g. $1$ device per 350 $m^{2}$,
we observe that communication services only support $19.5\%$ beyond
cellular coverage under the requirement of $90\%$ because of only
$3$ neighbors at each hop available to help. On the other hand, without
geo-controlled NC functionality, the performance degrades dramatically
with respect to the increase of distance. Transmission range is then
only extended approximately $9.5\%$ beyond cellular coverage constrained
by $90\%$ reliability in case of 1 device per 350 $m^{2}$. Specifically,
simulation results show that our proposed solution can provide up
to $18$ times and $12$ times gain in connectivity if compared to
transmission without geo-NC with $95\%$ reliability in cases of high
and low device's densities, respectively. 

\section{Conclusions \label{sec:CONC}}

In this paper, through analyzing the NC virtualized design in supporting
geo-controlled network reliability, we show the applicability of NCFV
in improving network reliability beyond cellular coverage where geographical
information is the key enabler to support network reliability. This
work is the first step towards the integration of NC virtualization
into future network systems including cooperative systems with heterogeneous
network devices.

\section*{Acknowlegment}

This research was financially supported by H2020 GEO-VISION - GNSS
driven EO and Verifiable Image and Sensor Integration for mission-critical
Operational Networks (project reference 641451). 

\bibliographystyle{IEEEtran}
\bibliography{NCFV}

\end{document}